# Atomic scale understanding of initial Cu-Ni oxidation from machine-learning accelerated first-principles simulations and *in situ* TEM experiments

Pandu Wisesa[1], Meng Li[2], Matthew T. Curnan[1,2,3], Jeong Woo Han[3], Judith C. Yang[2,4,5], and Wissam A. Saidi[1]*

[1] Department of Mechanical Engineering & Materials Science, University of Pittsburgh, Pittsburgh, PA 15261, USA

[2] Department of Chemical and Petroleum Engineering, University of Pittsburgh, Pittsburgh, PA 15261, USA

[3] Department of Chemical Engineering, Pohang University of Science and Technology (POSTECH), Pohang, Gyeongbuk 37673, Republic of Korea

[4] Environmental TEM Catalysis Consortium (ECC), University of Pittsburgh, Pittsburgh, PA 15261, USA

[5] Department of Physics, University of Pittsburgh, Pittsburgh, PA 15261, USA

* Correspondence to alsaidi@pitt.edu

**Abstract**

The development of accurate methods for determining how alloy surfaces spontaneously restructure under reactive and corrosive environments is a key, long-standing, grand challenge in materials science. Current oxidation models, such as Cabrera-Mott, are based on macroscopic empirical knowledge that lacks fundamental insight at the atomic level. Using machine learning-accelerated density functional theory with *in situ* environmental transmission electron microscopy (ETEM), we examine the interplay between surface reconstructions and preferential segregation tendencies of CuNi(100) surfaces under oxidation conditions. Our modeling approach based on molecular dynamics and grand canonical Monte Carlo simulations shows that oxygen-induced Ni segregation in CuNi alloy favors Cu(100)-O c(2×2) reconstruction and destabilizes the Cu(100)-O (2√2 × √2)$R$45° missing row reconstruction. The underpinnings of these stabilization tendencies are rationalized based on the similar atomic coordination and bond lengths in NiO rock salt and Cu(100)-O c(2×2) structures. In situ ETEM experiments show Ni segregation followed by NiO nucleation and growth in regions without MRR, with secondary nucleation and growth of $Cu_2O$ in MRR regions. This further corroborates the simulated surface oxidation and segregation modelling outcomes. Our findings are general and are expected to extend to other alloy systems.

## Introduction

Understanding oxide growth mechanisms during metal oxidation is essential for preventing oxidation-induced corrosion[1-2], controlling oxide processing[3], and predicting catalytic reactions[4-6]. Classical metal oxidation theories, such as the Cabrera-Mott model[7], describe oxide growth processes as metal or oxygen diffusion through oxide films, assuming formation of uniform oxide films on metal surfaces. However, this assumption is proven to be incorrect for initial oxidation of many metals and alloys, such as Cu and Ni-Cr,[1, 8-9] on which oxide islands form instead of continuous oxide films. Oxide islands are usually epitaxial on metal substrates[9-10] and strongly resemble heteroepitaxial interfaces during deposition. The Stranski-Krastanov mechanism is used to explain the growth of epitaxial oxide islands, predicting multilayer growth dynamics along substrate surfaces[11-12]. However, a recent in situ ETEM experiment on $Cu_2O$ growth suggests a new mechanism for epitaxial oxide island growth that features unusual layer-by-layer growth along a preferred oxide facet, regardless of substrate orientation.[13] These findings underscore the complex nature of metal oxidation and the necessity of an atomic level understanding of the oxidation process.

Oxidation kinetics of alloys are further complicated by their multiple constituent elements, which may have different oxygen affinities. Namely, alloys feature synergistic or competitive surface and oxygen-induced segregation tendencies, which can generate regions that are enriched by or depleted of particular metal element concentrations. This occurs in binary alloys such as Al-Mg[14-16], Cu-Ni[17-19], Cu-Pt[17], Pb-Ni[20], Pb-Bi[20], or even more complex ones such as Ag-[Fe, Co, Ni][21] and FeCoNiCuPt high entropy alloys[22]. For metal elements that enrich surfaces and favor oxide formation, such as Mg in Al-Mg[14-15], no further reactions compete during oxidation. However, for alloys including Cu-Ni[17-19], Cu enriches surfaces while Ni favorably forms oxides. Multiple processes can occur in parallel, such as Ni segregation to surfaces, Cu movement to bulk, and surface metal atom oxidation. The interconnectivity of such different mechanisms convolutes investigations of granular alloy oxidation.

Previous studies on early-stage Cu(100) surface oxidation indicate different surface reconstructions are affected by oxygen partial pressure and temperature variations[23-26]. Experimentally observed interfaces include clean unreconstructed (100) surfaces, (100) surfaces with 25% O-occupied hollow sites, c(2×2), and Cu(100)-O $(2\sqrt{2} \times \sqrt{2})R45°$ missing row reconstructions (MRRs). MRRs are most frequently observed prior to $Cu_2O$ island formation[26]. Given Cu-Ni alloys have Cu-rich surfaces, their reconstructions can be surmised to feature additional kinetic complexities influencing Ni surface segregation and subsequent oxidation, especially at higher oxygen coverages. Hence, accounting for reconstructions at the initial stage of Cu-Ni surface oxidation is crucial to having a full understanding of nucleation and growth processes.

Our article focuses on early oxidation stages of Cu-Ni alloys, which are well-studied as model systems forming ideal mixtures with competitive oxidation behaviors. Cu and Ni are completely miscible down to ~300 °C[27], but $Cu_2O$ and NiO show very limited solubility in metallic phases. Standard Gibbs formation energetics favor NiO over $Cu_2O$. Depending on oxygen partial pressure ($p_{O2}$), at least one alloy component will oxidize, enabling systematic determination of how composition and phase transition affect competitive nucleation and oxidation. Pilling and Bedworth investigated Cu-Ni oxidation by thermal gravimetric analysis and found parabolic oxide growth rates vs. relative oxygen concentrations[28], but these investigations did not provide structural information. Wagner described Cu-Ni as one example of competitive oxide nucleation[29], albeit without structural information and only considering oxidation with pure oxygen. Several surface science studies examined oxygen interaction and strain-induced NiO nucleation on CuNi but did not discuss very early oxidation stages[30-32]. *Ex situ* TEM studies of Cu-Ni alloy oxidation reveal both Cu and Ni oxides form[33], but without temporal information needed to elucidate nucleation kinetics. In summary, while characterizing binary alloy oxidation is important to understanding oxidation behaviors in engineered materials[34], literature on early oxidation stages of Cu-Ni and other alloys is limited by comprehensiveness of analyses, especially within single works.

Herein, we computationally and experimentally characterize Cu-Ni structure and composition during oxidation, including early and late reaction stages while simultaneously characterizing interfacial energetics, dynamics, and kinetics. Specifically, we employ first-principles density functional theory (DFT) calculations to examine the interplay between Ni segregation energetics on Cu(100) surfaces and variously composed surface reconstructions. To overcome DFT limitations regarding connections across scales in length and time domains, we develop a deep neural network potential (DNP) to investigate thermodynamics and kinetics during Ni segregation for large system sizes, finite temperatures, and time scales. Using molecular dynamics (MD) and Monte Carlo (MC) approaches, Ni concentration trends are observed over Cu-Ni interfacial layers featuring different Cu(100) reconstructions at various extents of oxidation. Starting from clean Cu/Cu-Ni surfaces, interfacial reconstruction and evolution are modeled over early-to-late-stage oxidation using MD and grand canonical Monte Carlo (GCMC). Theoretical outcomes are reconciled using *in situ* environmental transmission electron microscopy (ETEM) experiments, combining disparate computational components into a holistic description of experimental Cu-Ni segregation and oxidation.

## Results and Discussion

We first investigate Ni segregation trends on pristine Cu(100) surfaces. Segregation energy is defined as the energy difference between an interfacial configuration with an impurity atom in one of its surface layers, and that same interface with a matching impurity in its bulk-like layers. Negative segregation energies indicate impurities are more stable on surface than bulk-like layers. Comparatively, a recent study by Garza *et al.* simulated Ni segregation on Cu(100) with an embedded-atom model (EAM) potential developed by Fischer *et al.*[35], and similarly reported that Ni is not favored on the Cu(100) surface but is more favored in the bulk layers.[19] Notwithstanding this difference, current results agree with pristine surface energies, which favor Cu ($0.092$ eV/Å$^2$) over Ni ($0.138$ eV/Å$^2$) surfaces.[36] Ni segregation is expected to occur on other low-index surfaces, given their lower surface energies.

Under oxidation environments, the stability and segregation preferences of Ni doped Cu(100) surfaces are expected to differ from those in vacuum. Ni interacts more strongly with oxygen than Cu, as inferred from exemplary Ellingham diagrams showing NiO to be more stable than CuO. Thus, Ni segregation in oxygen environments is determined via competition between surface energies not favoring, and oxygen interactions preferring, Ni segregation. One could surmise that as concentrations of oxygen atoms interacting with surfaces increase, Ni segregation tendencies will improve. However, Cu(100) transforms into different surface reconstructions with increased oxygen concentrations, including c(2×2) and MRR, before forming $Cu_2O$ islands.[26] Given the interactions between Cu surface reconstructions, Ni surface segregation propensity, and relative Ni and Cu oxygen affinities, oxidation outcomes of the Cu-Ni alloy system under particular temperature and oxygen environments are unclear.

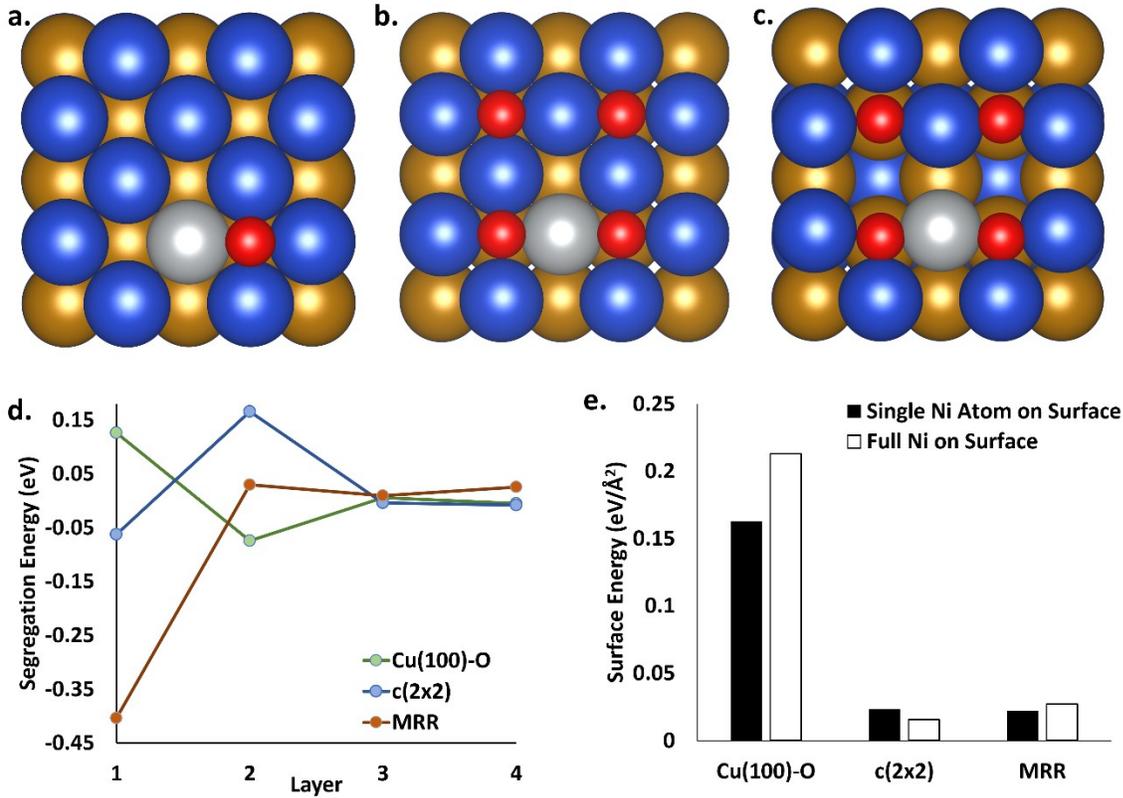

*Figure 1. Atomic models visualizing different surface terminations for **(a)** unreconstructed (100), **(b)** c(2×2) , and **(c)** MRR. Color-coded spheres represent Ni (silver) and oxygen (red), with Cu in odd (blue) and even (gold) layers. **(d)** Plotted segregation energies with oxygen on the surface vs. Ni layer position. **(e)** Surface energies for different surface terminations with single Ni on surfaces (black bars) and all surface sites filled with Ni (white bars).*

The interplay between Ni segregation and oxygen-induced Cu surface reconstructions was investigated first with DFT calculations by employing three different 3x3 supercell surface slab models (dilute limit = 1/18 O site coverage). This includes Cu(100) with single surface O, c(2×2), and MRR, as seen in **Figure 1a-c**. For simplicity, only single Ni substitutionally doped surfaces are considered. **Figure 1.** Atomic models visualizing different surface terminations for **(a)** unreconstructed (100), **(b)** c(2×2) , and **(c)** MRR. Color-coded spheres represent Ni (silver) and oxygen (red), with Cu in odd (blue) and even (gold) layers. **(d)** Plotted segregation energies with oxygen on the surface vs. Ni layer position. **(e)** Surface energies for different surface terminations with single Ni on surfaces (black bars) and all surface sites filled with Ni (white bars).

 confirms that low surface oxygen concentrations favor Ni occupation of subsurface sites, similar to pristine (100) surfaces  except that Ni is near O. Benefiting from Ni-O surface interactions, segregation energies of Ni in the presence of oxygen diminish by ~0.2 eV/Ni, or from 0.33 to 0.12 eV/Ni on respective pristine and oxidized surfaces (**Figure 1d**). While Ni is still unfavorable on top of surfaces, this significant decrease in segregation energy highlights the effect that oxidation has on Cu-Ni alloys, suggesting higher oxygen concentrations will inevitably induce Ni segregation to the surface.

Oxygen-decorated Cu(100) reconstructs into c(2×2) first, then MRRs with higher oxygen coverages.[26] Thus, we then examined the change in thermodynamic stability of these surfaces with the inclusion of single Ni atoms by exploring all symmetrically distinct sites (**Figure 1b-c**), surveying each surface, sub-surface, and bulk-like layer. The MRR is found to have two distinct substitutional sites for Ni on the surface – next to the missing row (segregation energy = -0.40 eV/Ni, **Figure 1c**) and between fully filled rows (segregation energy = -0.33 eV/Ni). As anticipated, the segregation energy in **Figure 1d** shifts from positive for the low oxygen coverage Cu(100)-O to negative for the higher coverage c(2×2), decreasing even more so for the highest oxygen concentration of the three (MRR). The difference between segregation energies of c(2×2) and MRR subsurface layers is rationalized due to the nature of the broken bonds between the rows in the MRR, allowing for easier movement of Ni atoms approaching bulk layers. In contrast, c(2×2) is fully occupied with oxygen bonded to multiple Cu atoms. For all surface models including the unreconstructed surface, **Figure 1d** shows that the segregation energies converge to the bulk value of zero at deeper layers. This is justified given that the bulk layers are all Cu, as they are far from the oxygen-decorated surface.

Trends in surface energies demonstrate the relative favorability of different Cu-Ni surface terminations and Ni concentrations, as shown in **Figure 1e** (black bars). Owing to this, Ni can be projected to eventually segregate towards the surface in the thermodynamic limit. Error! Reference source not found.**e** (white bars) shows surface energies for all terminations when all the surface Cu atoms have been substituted with Ni. Interestingly, all terminations except c(2×2) exhibit an increase in surface energy. This is likely due to the formation of relatively stable Ni-O, versus existing Cu-O, bonds. Moreover, atoms located alongside missing rows are undercoordinated, thus MRRs produce broken bonds that weaken hypothetically forming surface oxides. Resultantly, c(2×2) interfaces are expected to be more stable than MRRs as segregated Ni surface concentration increases.

While DFT simulations can provide insights on energetics for the two limiting cases of single Ni or full layer Ni substitutions on the surface, investigations of the full segregation dynamics and possible configurations of the CuNi(100)-O system are not computationally feasible with this approach. To address this challenge, we develop a DNP for the system building on configurations generated in previous studies[26, 37-39]. We posit that the DNP approach can address the complexities in these systems, given the successes of DNPs in robustly replicating DFT values across single elements[40-41], bimetallic systems[16, 42-43], supported metal nanoclusters[44], hybrid perovskites[45], and metal oxides[46]. The developed Cu-Ni-O DNPs are able to replicate Cu-Ni surface trends obtained before[19] under vacuum conditions, wherein Ni preferentially avoided surface layers. However, consistent with current DFT values and contrasting with results[19] obtained using a classical interatomic potential, DNPs slightly favored sub-surface Ni over bulk-like layers.

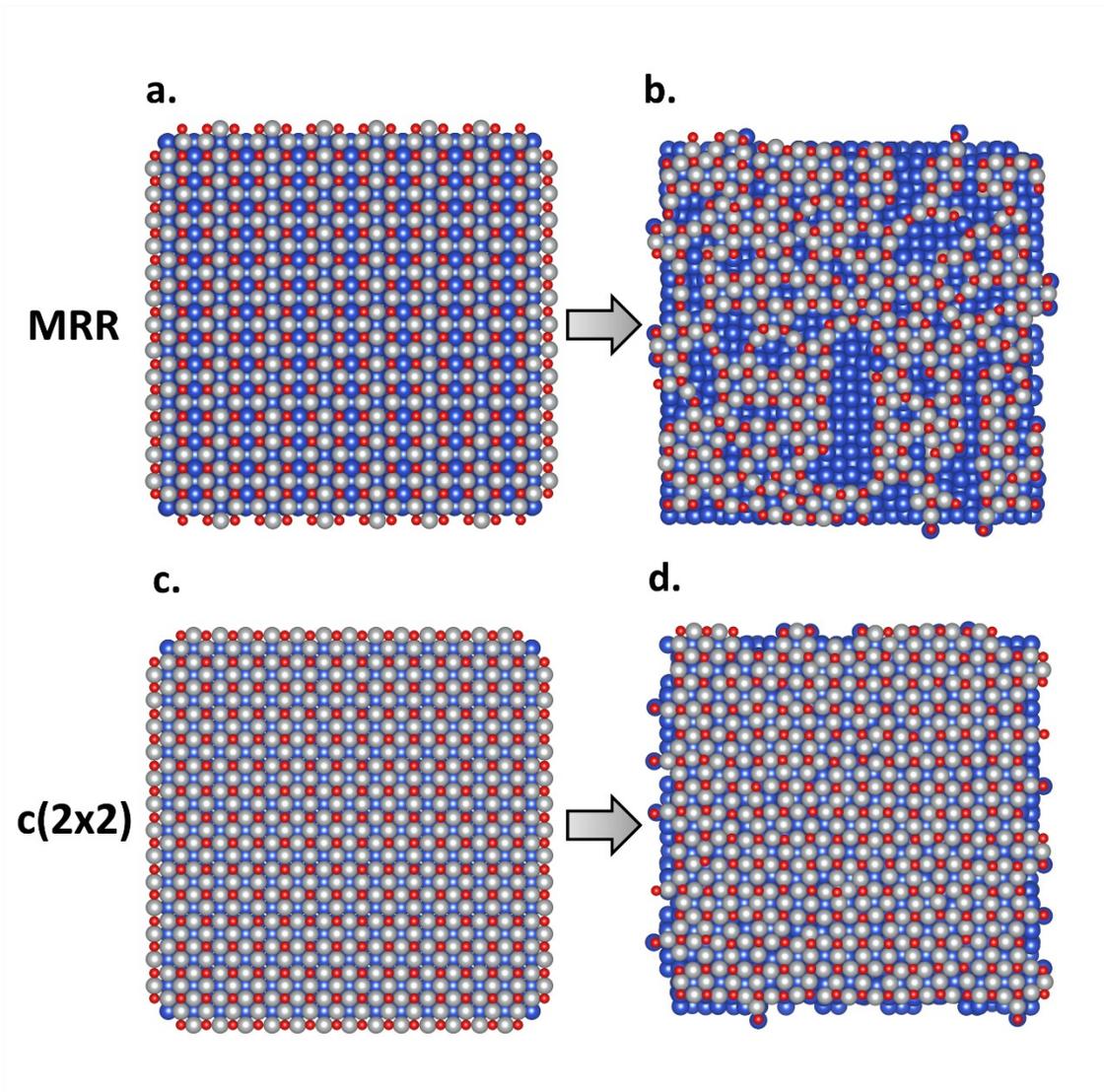

**Figure 2**. *MD simulation snapshots of faces for **(a,b)** MRR and **(c,d)** c(2×2) terminations during their **(a,c)** initial and **(b,d)** time-dependent states, with surface layer Ni (silver) substituting Cu (blue) atoms. Both Ni and oxygen (red) undergo significant movements within the MRR surface, but are stable within c(2×2).*

Using the developed DNP, surface evolution is examined over time on MRR and c(2×2) interfaces featuring surface Cu completely substituted with Ni. Simulations were run under a constant atom number, volume, and temperature ensemble (NVT) using 6x6 supercells at 800 K. They were propagated for 1 nanosecond (ns) at 1 femtosecond (fs) per step, taking snapshots every 1000 steps. **Figure 2** shows the initial and final configurations for the two surfaces. Ni and O in MRR terminations readily migrate at the early stages, creating nickel oxide islands and exposing sub-surface Cu. Thus, MRR configurations consisting of Ni and O are unstable. Contrastingly, Ni and O on c(2×2) maintain their positions throughout simulations, indicating Ni-O bond stability. These results are consistent with surface energetics from DFT favoring Ni segregation on c(2×2), and will also be vindicated by ETEM images presented later.

MD simulations sufficiently demonstrate atomic movement on single layers. However, physically realistic atoms also move between layers, though at much slower time scales. A recent study on CuNi carried out using accelerated MD and kinetic Monte Carlo indicated corresponding Ni migration time scales of around $\sim$1 s, which are inaccessible using MD simulations even after acceleration with DNPs.[19] To elucidate the instability during CuNi interlayer diffusion, Ni concentration persistency within MRR layers is explored while initializing Ni on exclusively surface or only random layers. To this end, we used a hybrid Monte Carlo (MC) and MD approach, which stochastically globally optimizes over configurations to ideally resolve equilibrated interfaces. MC/MD also advantageously accesses particular configurations to evaluate their favorability, unrestricted by kinetic requirements. The main drawback of stochastic approaches, including MC, is that they require sampling of many configurations ($\sim 10^5$ or more) to sample equilibrium profiles, which is possible in our study via the developed DNP.

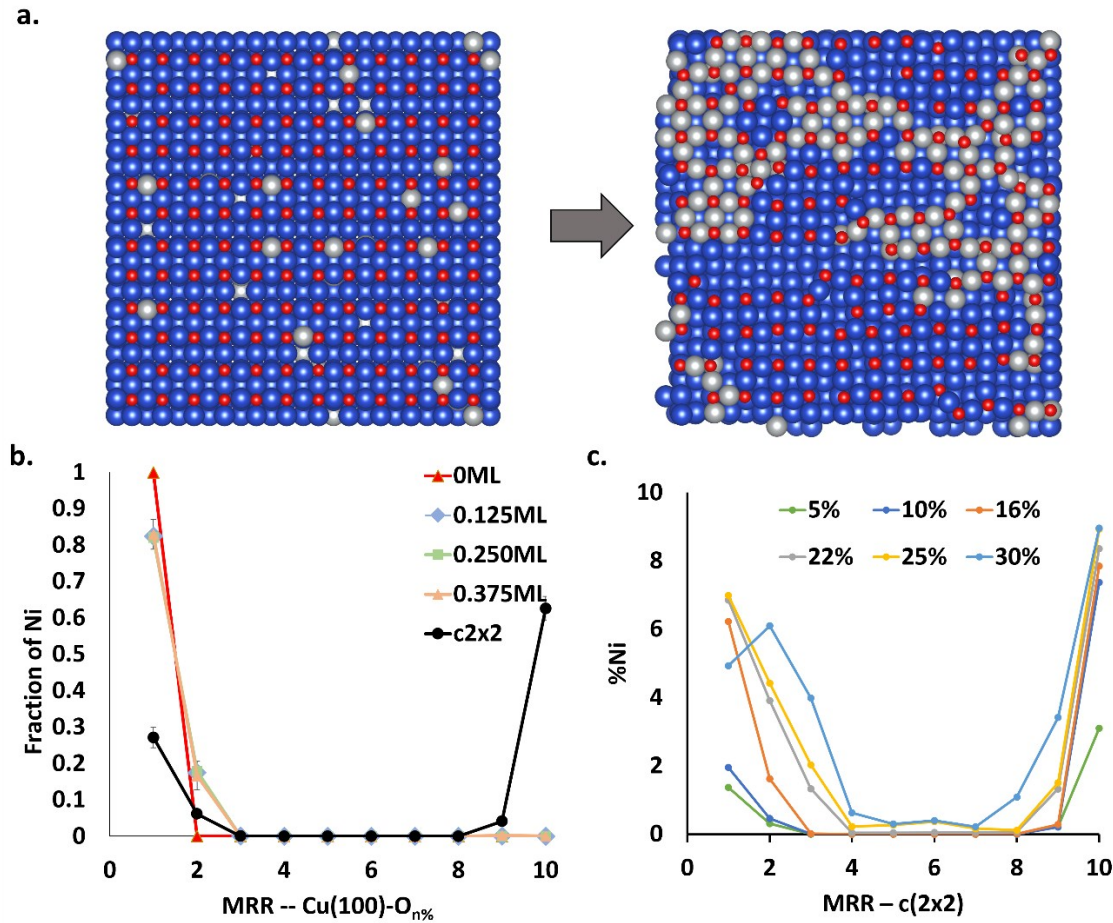

*Figure 3. Initial and time-dependent states of MRRs with 5% Ni random site substitution shown from the (a) top-down view of MRR faces. (b) Per-layer distributions of 5% total Ni substituted Cu sites across 6x6 supercell MRR-(100) double-sided slab models at various extents of surface oxidation; layer 1 is a MRR and layer 10 is Cu(100) with varying O concentration. (c) Distributions of substituted Ni percentages in each layer for various total substituted Ni percentages; layer 1 is a MRR and layer 10 is c(2×2).*

**Figure 3** summarizes our results obtained using the MC/MD hybrid approach. The simulations were run at 500 K for 1 million steps, wherein Ni and Cu may be exchanged every 20 steps, while NVT MD runs every step at a 2 fs time-step. We used an alloy with randomly placed 10% Ni on ordered MRR terminations. **Figure 3a** shows the initial structure and a configuration from the MC/MD ensemble after reaching equilibrium. As surfaces evolve over time, more Ni migrates to surfaces, given attraction to oxygen. In conjunction, oxygen atoms are attracted to Ni, leaving their initial positions to bond with Ni, clustering together to form islands, and reconstructing surfaces simultaneously. The configuration from the MC ensemble shown in **Figure 3a** emphasizes the MRR instability in oxidized Cu-Ni systems, and suggests that NiO-rich and Cu-rich islands are likely to form separately in experimental studies. This also can be seen form the layer-by-layer averaged concentration profiles obtained from the MC ensemble for the 9-layer slabs, wherein all Ni has segregated to the MRR surface, as seen in **Figure 3b** (red line).

In experimental environments featuring ambient oxygen pressure and compositionally homogenous alloy surfaces, uniform oxygen coverage over surfaces can be realistically assumed. Hence, comparisons between MRRs and surfaces with lower discretized oxygen coverages are not entirely physical. Nonetheless, these comparisons determine whether studied reconstructions commonly occur. Namely, whether Ni prefers migration towards surface areas with different surface composition, and how surface oxygen concentration influences Ni segregation favorability. To further investigate this, CuNi slab models are employed with a 5% Ni concentration, in which one (top) surface of the slab (layer 1) exposes a MRR and the bottom one (layer 10) exposes oxygen-decorated Cu(100) or a c(2×2) reconstruction. The simulation of **Figure 3a**, wherein the MRR is on the top surface and Cu(100) is on the bottom surface, can be regarded as a special case of oxygen-decorated Cu(100) with 0 ML oxygen coverage. **Figure 3c** shows layer-by-layer averaged Ni concentration profiles obtained from the MC ensemble for different oxygen coverages on unreconstructed Cu(100) surfaces and c(2×2). As seen from the figure, MRRs are strongly favored by Ni versus Cu(100) surfaces with lower oxygen concentrations, as almost all of the Ni segregated to the MRR surface (layer 1). This changes in MRR versus c(2×2) comparisons, where ~65% of all Ni favor c(2×2) (layer 10) terminations. Correspondingly, ~27% of Ni segregate to the MRR (layer 1), forming islands as demonstrated previously.

Under conditions better resembling experimental surface oxidation, surplus Ni supply would enable continuous surface segregation within energetically favorability limits. These thresholds represent how much Ni can be accepted by given surface terminations, after which Ni will segregate to different layers or surface terminations. **Figure 3c** represents substituted Ni concentrations as percentages in individual layers, relative to total substituted Ni percentages, that cumulatively sum to total numbers of sites over all layers. This allows direct comparisons between different total Ni concentrations. For reference on changes in trends, the 5% (green) line is the same as that for the MRR versus c(2×2) comparison in **Figure 3b**. As seen in the figure, Ni segregation consistently favors c(2×2) slab surfaces, which contain ~75% of Ni substitutions when 10% of all Cu sites are Ni substituted (~7% of cumulative sum). With 30% of total Ni sites substituted, c(2×2) layers feature ~30% of substituted Ni sites (~9% of cumulative sum), implying near c(2×2) surface layer saturation. Despite large total Ni concentration improvement, Ni concentration in c(2×2) surface layers only slightly increases. This evidences that surface saturation, when also considering c(2×2) sub-surface layers, is overall more occupied by Ni than at lower total Ni concentrations. When total Ni concentration increases from $10-16\%$, cumulative sums of Ni concentration in c(2×2) sub-surface layers increment from ~$0.2-0.3\%$, ultimately featuring ~50% Ni substitution (~8% of cumulative sum) on c(2×2) surface layers. This threshold approximates the Ni saturation point for c(2×2) surfaces.

On MRR surface sides, additional complexities arise when estimating per-layer Ni concentrations as surfaces appreciably reconstruct, as shown in **Figure 3a**. Reconstruction generally increases Ni segregation favorability, particularly up to 28% Ni (7% of cumulative sum) on MRR surfaces at 25% total Ni. As total Ni concentration improves to 30%, MRR surface concentration falls to 16% (5% of cumulative sum), as MRR sub-surface layers favor Ni segregation more ( 20% Ni, 6% of cumulative sum) and further underlying layers incorporate more Ni. These changes portray complexities in how interlayer diffusion affects Ni segregation favorability. For certain surfaces, Ni segregation tendencies highly depend on O concentration, as all terminations with lower O coverage than MRRs prefer Ni less. MRRs themselves are not as preferred as c(2×2), though they still attract Ni because MRRs reconstruct into islands resembling localized c(2×2). Thus, Ni readily segregates to surfaces, given sufficient oxygen exists to form c(2×2). Ni will not form MRR terminations, but rather attract oxygen to form NiO and c(2×2). When oxygen concentration is insufficient to form NiO, c(2×2) will develop instead.

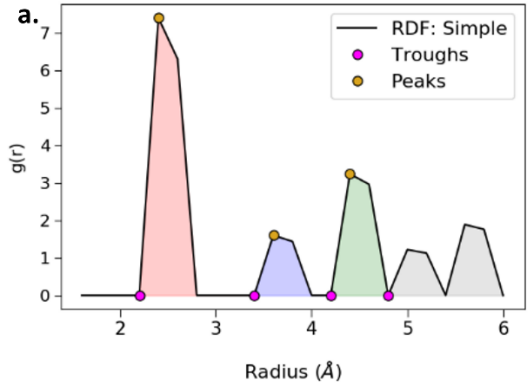
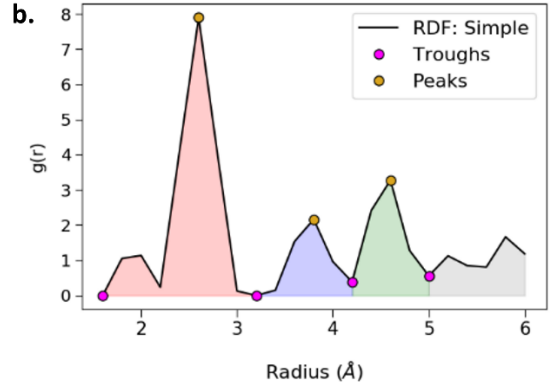
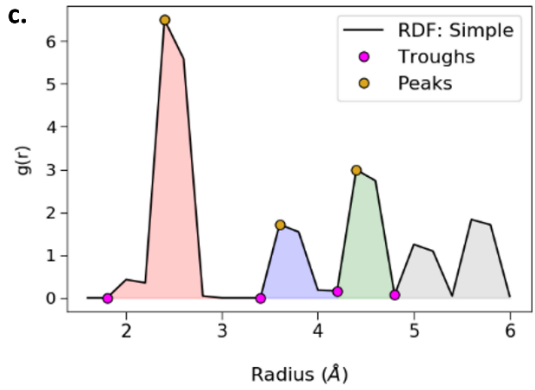
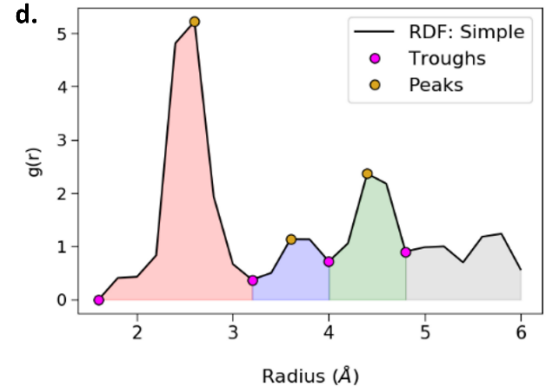
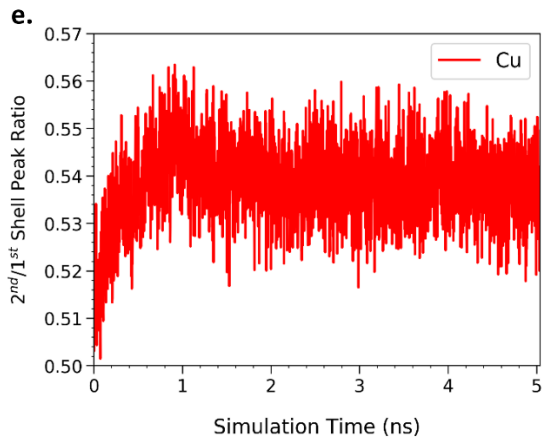
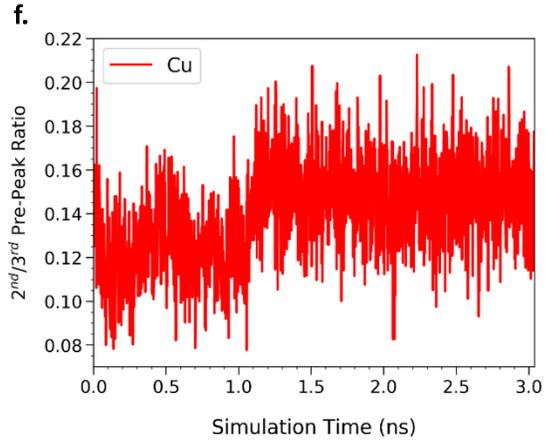
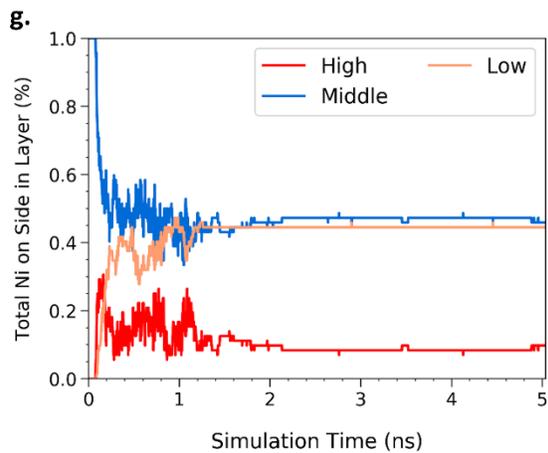
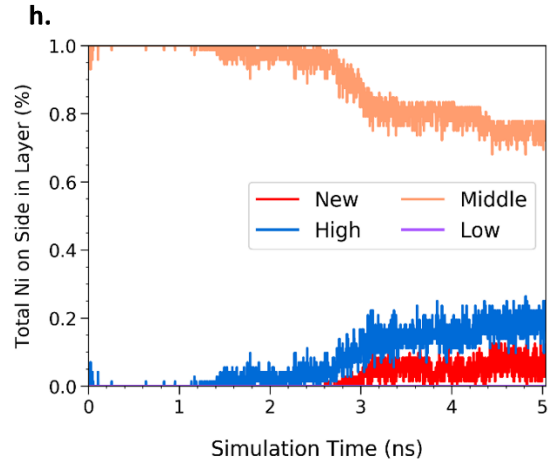

**Figure 4.** *RDFs of* **(a)** *Cu(100),* **(b)** *c(2×2), and* **(c)** *MRR interfaces relaxed at 0 K, and* **(d)** *c(2×2) from MD/GCMC at 300 °C. The 1st (red), 2nd (blue), and 3rd (green) common coordination shells over all structures are filled according to color-coded shading, while shell peaks (gold) and troughs (purple) are marked accordingly.* **(e)** *Ratios between peak magnitudes of 2nd and 1st coordination shells, coupled with* **(f)** *those of second (2.0 Å) and third (2.2 Å) binned RDF distances associated with reconstruction pre-peaks, measured throughout Cu(100) MD/GCMC surface reconstruction.* **(g)** *Percentages (%) of Ni atoms (normalized relative to initialized single middle layer totals) on bottom (nearest bulk), middle, top (uppermost layer before oxidation), and newly reconstructed surface layers of non-oxidized and* **(h)** *oxidized CuNi(100) throughout MD/GCMC at 400 °C.*

**Figure 4** respectively characterizes oxide growth and Ni segregation initialized on pristine Cu(100) and CuNi(100) surfaces through hybridized MD/GCMC simulations, which first relax and thermalize surfaces via MD. Subsequently, these surfaces are oxidized via GCMC for 1 ns to equilibrate concentrations of O atoms on them. Resulting interfaces are then structurally equilibrated via MD for another 2-4 ns, reconstructing Cu surfaces and segregating Ni from subsurface layers towards either surface or bulk. Expected radial distribution functions (RDFs) of Cu surface reconstructions, generated from 0 K atomic models, are displayed in **Figure 4a-d.** Cu(100) RDFs portray three metal-based characteristic peaks at approximately 2.6, 3.6, and 4.4 Å, which correspond to the 1st, 2nd, and 3rd Cu coordination shells. They respectively represent the 1st, 2nd, and 3rd nearest neighbor (NN) bonding of all Cu atoms. By extension, c(2×2) reconstructions have added broad and smaller peaks encompassing 1.8-2.2 Å, which depict idiosyncratic Cu-O surface bonds (1.96 Å). With further perturbation, these peaks preceding metal-based shells ("pre-peaks") drop less abruptly between 2.0-2.2 Å, effectively broadening and merging with 1st NN peaks of diminished magnitudes. In combination with flat troughs between 2.8-3.4 Å indicating atomic interaction absence, these RDF characteristics represent Cu MRRs. However, all of these RDFs depict structures at 0 K, while dynamical structures achieved above 0 K do not have RDFs with well-defined troughs constituting interaction absence. Given these considerations, ratios between the magnitudes of 1st coordination shell peaks, broadened pre-peak features, and reference peaks are calculated to discern reconstruction and Ni segregation over MD/GCMC simulation time.

**Figure 4e-f** quantitatively confirms Cu surface reconstruction via MD/GCMC dynamics, plotting ratios between 2nd and 1st shell peak magnitudes, as well as between pre-peak termination *g(r)* (*r*=2.2 Å) and 1st shells peaks, over simulation time. The structural transformation between Cu(100) and c(2×2) is evidenced by the inflection in 2nd vs. 1st shell peak ratios at ~1 ns and is verified using atomic models encompassing that time. Further, variation in pre-peak termination versus 1st shell peak ratios is detectable between 1-1.5 ns, substantiating a c(2×2) to MRR reconstruction that is also confirmed via atomic models.

To model Ni segregation on CuNi(100) surfaces, double-sided slabs were initiated with single Ni subsurface layers on both of their sides. Over these slabs, featuring three unfixed surface layers on each side and three fixed bulk layers in between terminal sides, these initial Ni rows are designated as the middle unfixed original surface layers. Each slab side also has corresponding "high" and "low" layers encompassing these "middle" layers, while oxidized sides feature "new" layers that result from "high" row reconstruction and Ni segregation. **Figure 4g-h** confirms Ni segregates mostly to bulk or "low" layers on non-oxidized sides throughout MD/GCMC simulation time. Conversely, oxidation does not initially induce Ni segregation towards surfaces via GCMC. Nevertheless, reconstruction of CuNi surfaces throughout long MD simulations following GCMC induces Ni segregation more towards surfaces, as well as no significant Ni diffusion towards bulk. Therefore, oxidation determines Ni segregation direction in CuNi interfaces, consistent with past simulations (**Figures 1-3**) and experiments (**Figure 5**).

In situ Environmental TEM (ETEM) experiments were performed on CuNi alloys during oxidation to verify our predictions. 60 nm thick Cu-5at.% Ni(100) alloy thin films were first annealed in 0.01 Pa $H_2$ at 600 ˚C for 30 minutes inside the ETEM to reduce native oxides, yielding clean CuNi surfaces. To better observe early-stage surface reconstruction dynamics, in situ oxidation experiments were performed at a lower pressure of $7\times10^{-3}$ Pa $O_2$ under 300˚C. The captured experimental data were aligned at an atomic level using a customized algorithm introduced in our previous paper[47]. **Figure 5** shows snapshots of oxidation processes obtained via *in situ* ETEM, with **Figure 5a** confirming CuNi film surfaces were unreconstructed in vacuum before $O_2$ injection. After $O_2$ injection, the $O_2$ pressure gradually increased to its set point at ~60s. During the increase of $O_2$ pressure, c(2×2) reconstructions were first observed (**Figure 5b**). Considering Cu and Ni share the same crystal structure (FCC), close lattice constants (3.613 Å vs. 3.521 Å), and similar atomic numbers, discerning early-stage Ni segregation from ETEM images is difficult. However, after Ni segregate to a few atomic layers, the strain between Ni (unit cell lattice constant = 3.521 Å) and Cu (3.613 Å) lattices became visible as marked by the arrows in **Figure 5c**. Also, **Figure 5c** shows MRR are formed next to, rather than within, Ni-segregated regions. Consistent with **Figure 1e**, MRRs produce broken bonds inhibiting nearby surface Ni-based oxide formation, while c(2×2) surface energy becomes more favorable when adjacent to stable Ni-O bonds encountered during Ni segregation. Therefore, MRRs only develop experimentally after initially homogenously dispersed Ni segregate from them to further stabilize c(2×2). Later, these Ni segregation regions grew into epitaxial NiO nano-islands (**Figure 5d**), mirroring the spontaneous migration of Ni away from MRRs observed in prior MD simulations (**Figure 2**). Furthering computational and experimental agreement, Ni atoms near c(2×2) retain their positions owing to stable bonds with O, forming NiO islands in MD and under ETEM. During NiO nano-island growth, **Figure**

**5d** shows that MRRs remain on Ni-free surface sections. After NiO nucleation and growth, $Cu_2O$ secondarily nucleates and grows between NiO islands on MRR surfaces (**Figure 5e**). During oxidation processes, MRRs are absent from Ni-segregated surface regions, which later grow NiO to confirm their high Ni concentrations. These observations are further corroborated by hybridized MD/MC outcomes from **Figure 3**, which initially depict c(2×2) formation and subsequently demonstrate NiO growth in similar regions as Ni concentration sufficiently increases. In comparison, $Cu_2O$ was observed on MRR regions, indicating these regions are Cu-rich. Reconciling MD (**Figure 2**), hybridized MD/MC (**Figure 3**), and ETEM findings (**Figure 5e**), Cu-rich MRR and Ni-segregated c(2×2) surface regions ultimately grow separate $Cu_2O$ and NiO island phases, respectively.

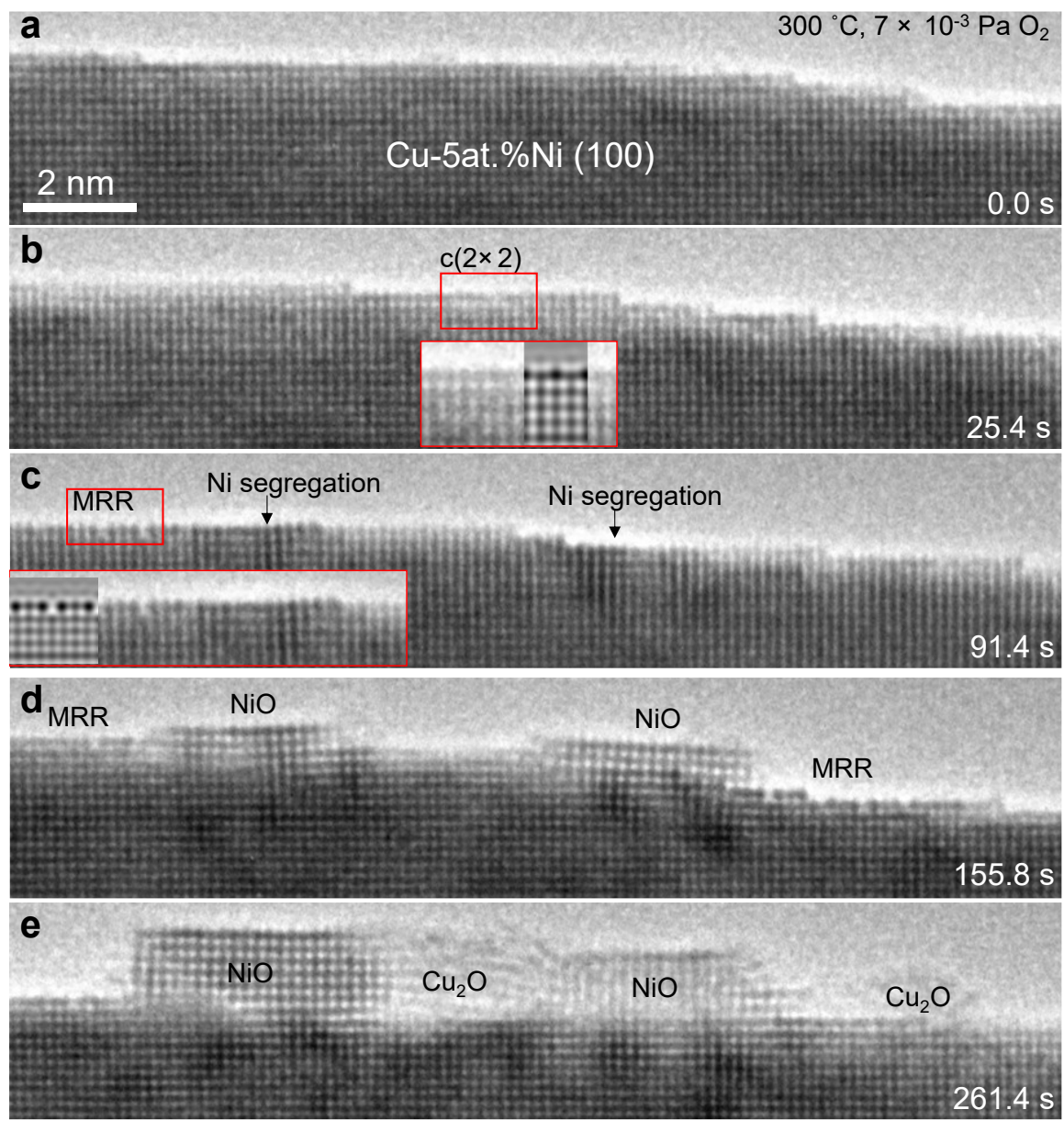

**Figure 5.** *In situ ETEM observation of Cu-Ni(100) (5 at.% Ni) thin films during early-stage oxidation.* **(a)** *Under vacuum (5x10⁻⁴ Pa), alloy surfaces were reconstruction-free.* **(b)** *During $O_2$ injection, $O_2$ partial pressure gradually increased and reached set pressure at ~60 s. c(2×2) reconstructions were first observed at lower $O_2$ pressure, as shown in the enlarged image (inset) and verified by HRTEM simulation overlayed on the enlarged image.* **(c)** *After reaching set gas pressure, MRRs started to develop in some surface regions, and Ni segregation was observed in regions without MRRs. Insets show enlarged views of MRR regions (right) and HRTEM simulations of MRRs (left).* **(d)** *NiO islands nucleated in Ni segregated regions. MRRs are observed to cover most alloyed surfaces.* **(e)** *Secondary nucleation of $Cu_2O$ islands was observed between NiO islands.*

**Conclusion**

In initial DFT simulations, CuNi(100) favors Ni migration to subsurface sites one layer below surface layers without O. Throughout the c(2×2) and MRR stages of surface reconstruction, Ni segregation to surfaces is respectively energetically favorable and unfavorable. MC simulations, completed using DNPs trained on previously detailed DFT results, corroborated first-principles outcomes. To achieve equilibrium, Ni and O atoms hopped from MRR sites towards c(2×2) sites during MC. Subsequent hybrid MD/MC simulations confirm Ni diffuses towards CuNi(100) surfaces in equilibrium, but only when sufficient surface O concentrations are present. Correlations between per-layer Ni and O concentration thresholds are then connected to different stages of Cu surface-based reconstructions, including c(2×2) and MRR. Cu and CuNi(100) surface evolution dynamics are compared via hybrid MD/GCMC simulations, which are related to referential RDFs resolved for structures at 0 K. Characteristic features involving RDF peaks are first generalized from idealized 0 K structures. Then, they are reconciled with atomic models and proportional RDF features observed during oxidized and non-oxidized surface evolution under reaction conditions. Overall, Ni selectively segregated towards oxidized CuNi(100) surfaces that were formerly c(2×2) reconstructions while Cu surfaces transformed into MRRs, consistent with past findings. Lastly, experimental *in situ* ETEM results vindicated simulation results depicting observed reconstructed phases, selective Ni segregation, and surface evolution over time. Furthermore, extended experimental time scales revealed the ultimate nucleation and growth of separate Cu and Ni oxide islands, which were developed in coordination with one another but occurred in sequential order.

These findings reveal how Ni alloying impacts Cu surface oxidation and resulting reconstruction over time. Thereby, design principles are developed for controlling corrosive, electrocatalytic, and related active site availability throughout oxidation reactions. Furthermore, integrated implementation of DFT, DNP optimization, MD, MC, GCMC, and other approaches in this work provides a template for correlating structural dynamics and thermodynamic outcomes, which can be adapted to future research studying interfacial evolution, corrosion prevention, and catalytic site activation.

**Methods**

All DFT calculations were done with the Vienna Ab initio Simulation Package (VASP)[48-49]. The Perdew-Burke-Ernzerhof (PBE)[50] exchange-correlation functional was used to solve the Kohn-Sham equations. Projector augmented wave (PAW)[51] pseudopotentials were applied to represent electron-nucleus interactions[52]. Planewave basis set cut-offs of 400 eV and $k$-grid densities of 0.24 Å$^{-1}$ were used for all DFT calculations, as previous studies generated currently applied DNP databases and accurately reproduced physical properties under these criteria[40]. Self-consistent electronic cycles terminated at tolerances of $10^{-8}$ eV.

Implemented DNPs were developed with the DeepPot-SE method,[53] as implemented in DeePMD-Kit[54]. Neighbor searching cut-off and smoothing function initialization radii of 7 and 2 Å were respectively imposed. Embedding and fitting nets of 25 x 50 x 100 and 240 x 240 x 240 were applied, respectively. Neural networks were trained using the Adam stochastic gradient descent method, initializing exponential moving average learning rates at 0.001. Loss function prefactors for energy, forces, and virials were kept at constant values of 1, 10000, and 10, respectively. All aforementioned hyperparameters were derived from previous studies[16, 40, 42-43]. DNP datasets were built according to structures from previous literature on Cu surface oxidation.[26, 37-39] Atomistic simulations applying DNPs were executed with LAMMPS[55].

*In situ* environmental transmission electron microscopy (ETEM) was used to verify theoretical predictions. 60 nm thick single-crystal Cu-Ni(100) thin films (5 at.% Ni) were prepared using Ultra-High Vacuum (UHV) *e*-beam evaporation on NaCl(100) substrates, and were then transferred to Cu TEM mesh grids using the float-off method introduced in our previous paper[13]. We used a differentially pumped ETEM (Hitachi H-9500, LaB$_6$) operated at 300 keV, equipped with a double-tilt heating holder (Hitachi) and a homemade gas injection system with up to 3 lines of gas injection (NFCF, University of Pittsburgh). Native oxides on CuNi films were removed by annealing in H$_2$ at an O$_2$ partial pressure ($p_{O2}$) of $1.0\times10^{-2}$ Pa and temperature ($T$) of 600 °C. Elevated temperature annealing also facilitates formation of faceted holes with (100) and (110) facets on CuNi films, enabling ensuing observations of oxidation from the edge-on viewpoint. Then, oxidation experiments were carried out under $p_{O2}$=$7\times10^{-3}$ Pa and $T$=300 °C. Real-time atomic-resolution movies of CuNi(100) facet oxidation were captured under HRTEM imaging using a Gatan Orius 833 CCD camera, with a frame rate of 5 frames per second.

## Acknowledgements

We are grateful to the U.S. National Science Foundation (Award No. CSSI-2003808). This research was supported in part by the University of Pittsburgh Center for Research Computing through the resources provided. Specifically, this work used the H2P cluster, which is supported by NSF award number OAC-2117681. GCMC simulations within this work were supported by the Basic Science Research Program (NRF-2022M3H4A1A01008918 and NRF-2021R1A2C3004019) offered through the National Research Foundation of Korea (NRF) grant funded by the Korean government (MSIT). The in situ ETEM experiment is supported by NSF CMMI-1905647, and used resources from the Petersen Institute of NanoScience and Engineering (PINSE) Nanoscale Fabrication and Characterization Facility (NFCF) at the University of Pittsburgh.